\documentclass[journal]{IEEEtran}

\usepackage[justification=centering]{caption}
\usepackage{graphicx}
\usepackage{subcaption}
\usepackage[english]{babel}
\usepackage[utf8x]{inputenc}
\usepackage{amsmath}
\usepackage{tikz}
\usepackage{xspace}
\usepackage{siunitx}
\usepackage{lettrine}
\usepackage{enumitem}
\usepackage{amsmath}
\usepackage{amsfonts}
\usepackage{amssymb}
\usepackage{breqn}
\usepackage{lipsum}
\usepackage{etoolbox}
\usepackage{multirow}
\usepackage{tabularx}
\usepackage{longtable}
\usepackage{siunitx}
\usepackage{dblfloatfix}
\usepackage{float}
\usepackage{xcolor}
\usepackage[font=footnotesize]{caption}

\DeclareCaptionLabelSeparator{periodspace}{.\quad}
\usepackage{cite}
\usepackage{multirow}
\usepackage{hyperref}
\usepackage{graphicx}
\usepackage{amsmath}
\usepackage{multicol}
\usepackage{mathtools}
\usepackage{tabularx}
\usepackage{longtable}
\usepackage{mdwlist}
\usepackage[linesnumbered,ruled,vlined]{algorithm2e}
\usepackage{algpseudocode}
\usepackage{balance}
\usepackage[square,numbers]{natbib}
\usepackage{booktabs}
\begin{document}
\captionsetup[figure]{labelsep=colon,name={Fig.}}

\title{Bayesian Ridge Regression Based Model to Predict Fault Location in HVdc Network}

\author{\IEEEauthorblockN{Timothy Flavin\IEEEauthorrefmark{2}, Thomas Steiner\IEEEauthorrefmark{2},
Bhaskar Mitra\IEEEauthorrefmark{3}\IEEEauthorrefmark{1} and 
Vidhyashree Nagaraju\IEEEauthorrefmark{2}} \\

\IEEEauthorblockA{\IEEEauthorrefmark{2}Tandy School of Computer Science, The University of Tulsa, Tulsa, OK 74104 USA}\\

\IEEEauthorblockA{\IEEEauthorrefmark{3}Pacific Northwest National Laboratory, Richland, WA 99354 USA}\\

\IEEEauthorblockA{\IEEEauthorrefmark{1}Corresponding author: Bhaskar Mitra (email: bhaskarmitra1991@gmail.com)}}

\maketitle

\begin{abstract}
This paper discusses a method for accurately estimating the fault location in multi-terminal High Voltage direct current (HVdc) transmission network using single ended current and voltage measurements. The post-fault voltage and current signatures are a function of multiple factors and thus accurately locating faults on a multi-terminal network is challenging. We discuss a novel data-driven Bayes Regression based method for accurately predicting fault locations. The sensitivity of the proposed algorithm to measurement noise, fault location, resistance and current limiting inductance are performed on a radial three-terminal MTdc network. The test system is designed in Power System Computer Aided Design (PSCAD)/Electromagnetic Transients including dc (EMTdc).
\end{abstract}

\begin{IEEEkeywords}
Bayes Ridge Regression, Fault location, HVdc, Modular multi-level converter, MTdc network
\end{IEEEkeywords}

\section{Introduction}
High Voltage Direct Current (HVdc) transmission corridors are widely used for long distance transmission. It has emerged as a leading contender compared to High Voltage Alternating Current (HVac) for bulk power off-shore and on-shore transmission. They are considered as a viable form of energy transportation for renewable energy transportation and interconnection of multiple grids of different frequencies. The Voltage Source Converter (VSC) design allows for the implementation of interconnected multi-terminal dc (MTdc) networks with bi-directional power flow. The Modular Multilevel Converter (MMC) has emerged as a popular choice for VSC-HVdc networks due to salient features of better scalability and higher operational efficiency \cite{mitra2018hvdc}. Accurate fault location for HVdc transmission networks are essential to reduce the downtown and improve reliability of the system. Accurate location of faults on the network is thus essential for repair and restoration process. 

Several fault location techniques in HVdc systems namely (1) spectral analysis; (2) travelling wave; (3) impedance method and (4) machine learning approaches have been discussed previously. Methods incorporating S-transform to to locate faults have been discussed in \cite{Kang}, although the results of the proposed method have not been verified with measurements noise. Travelling based methods are employed on a wide scale for accurate fault location using (1) single ended and (2) double ended measurements \cite{Ando1985}.

Double ended measurements provide higher accuracy of fault locations using synchronized current and voltage measurements. This method involves setting up and maintenance of robust communication networks and expensive time synchronized devices \cite{Ying-HongLin2004}. On the contrary single ended measurements are more convenient. They are cheap but tend to provide inaccurate results as devices do not have the capability to detect the reflected peak \cite{Xu2011}. Its performance is driven by fault resistance and other network parameters. The reflected surge waves are weaker thus making their detection difficult. Inability to capture the reflected surge wave peaks would provide provide inaccurate fault locations.

Some other methods of fault location has been proposed using digital signal processing, they require devices with high sampling frequency to achieve the desired results. Methods involving time-frequency analysis of the fault transients have been performed using wavelet transform \cite{Magnago1998}, implantation of such algorithms would require double ended synchronized measurements. Passive methods of fault locations have been suggested in \cite{Mohanty2016} and \cite{Mitra2021}, although their efficacy for a real-field example are yet to be demonstrated.


In this paper, we propose a single ended fault location technique using Bayesian ridge regression coupled with systematic application of various data pre-processing methods. The single-ended post-fault voltage and current measurements are used as input to the Bayesian ridge regression model. To validate the selection of BRR and sequence of pre-processing methods, combination of seven machine learning methods, five feature extraction, and two dimensionality reductions were applied with range of values for configurable parameters. Results suggested that Bayesian ridge model demonstrated a minimum of 50-70\% better predictive capability, while maintaining third ranking based on information theoretic measures.

The remainder of the paper is organised as follows: Section \ref{sec:TestSystem} describes the test system and generation of data. Section \ref{sec:Method} describes the Bayesian Regression methodology and other preprocessing steps. Section \ref{subsec:GOF} discusses the parameters required to assess the model performance. Section \ref{sec:Experiment} discusses the results and Section \ref{sec:Conclusion} concludes the paper with major findings.

\section{Test System and Data Generation} \label{sec:TestSystem}

\subsection{Test System Design}
The modeling of the three terminal MMC in this paper is based on the design suggested in \cite{Debnath2018}. The MMC model consists of 400 half bridge sub-modules (HBSM) per arm. Hybrid discretization and relaxation algorithms described in \cite{Debnath2018} are used to define the numerical stiffness in the differential algebraic equations. Details about the system parameters are provided in Table \ref{Table 1}.

\begin{table}[!ht]
\centering
\caption{System Parameters}\label{Table 1}
\begin{tabular}{|c|c|c|}
\hline
 & \textbf{Parameters} & \textbf{Value} \\ \hline
\multirow{5}{*}{\textbf{ac side}} & Voltage (L-L RMS) & 333 kV \\ \cline{2-3} 
 & Length of transmission line 1 \& 3 & 100 km \\ \cline{2-3} 
 & Length of transmission line 2 & 150 km \\ \cline{2-3} 
 & System Frequency & 60 Hz \\ \cline{2-3} 
 & Transmission line resistance & 0.03206 $\Omega$/km \\ \hline
\multirow{4}{*}{\textbf{dc side}} & Voltage (L-L) & 640 kV \\ \cline{2-3} 
 & Length of transmission line & 1000 km \\ \cline{2-3} 
 & Transmission line resistance & 0.03206 $\Omega$/km \\ \cline{2-3} 
 & MMC capacity & 1 GW  \\ \hline
\end{tabular}
\end{table}

\begin{figure}[!ht]
    \centering
    \includegraphics[width=0.5\textwidth]{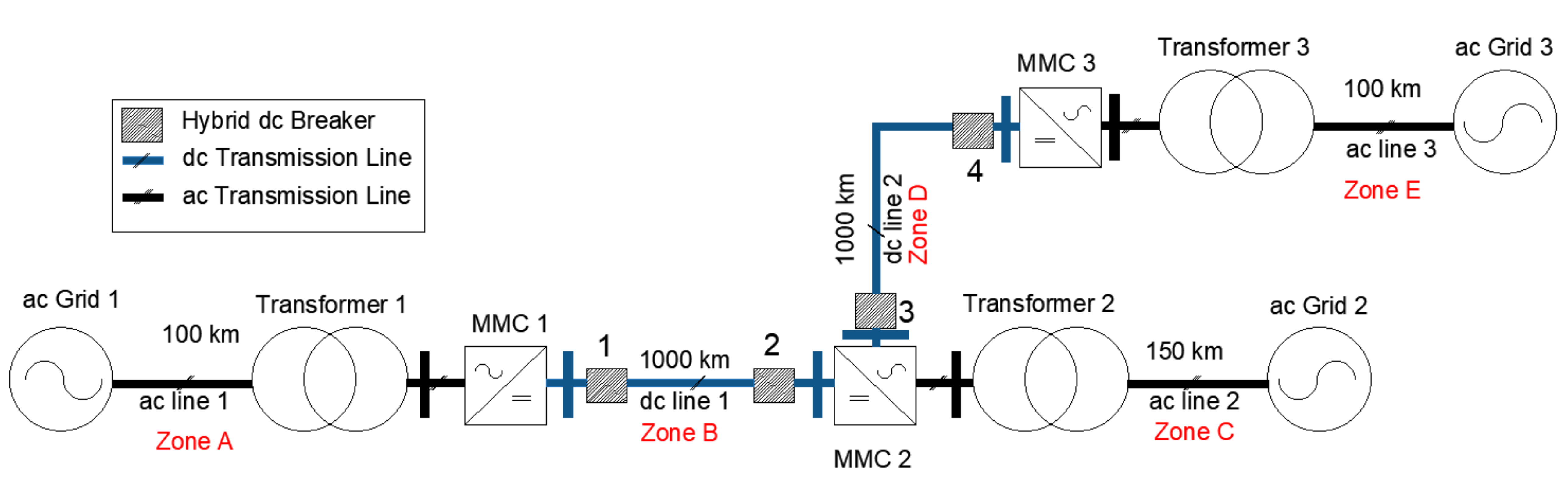}
    \caption{Radial multi-terminal dc network}
    \label{fig:zones}
\end{figure}

The model of a radial three-terminal MTdc symmetric monopole is shown in Fig.~\ref{fig:zones}. The system is equipped with hybrid dcCB at the MMC terminals. The dc transmission lines are designed as frequency dependent models having $6$ conductors with a vertical spacing of $5m$ and horizontal spacing of $10m$ between the conductors. 

\subsection{Data Description}\label{sec:Ex:ExperimentalSet}

As shown in Fig. \ref{fig:zones}, a 640 kV dc, radial MTdc transmission network was utilized for the single-ended fault location. The parameters for the test system are listed in Table \ref{Table 1}. Fault events were simulated on various sections of the dc transmission network using PSCAD/EMTdc. Faults were simulated at a separation of 25km across the network. Bus faults are typically rare and were not considered for this study. Scenarios were generated by varying the current limiting inductance (1mH - 200mH), fault resistance (0.01$\Omega$ - 200$\Omega$), and change of loads in the transmission system (non-fault events). The current and voltage at the terminals were sampled at $10kHz$.



\begin{figure*}[]
    \centering
    \subfloat[ Current inputs over time]{{\includegraphics[width=0.49\textwidth]{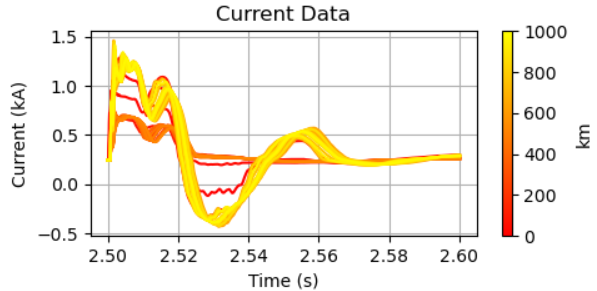} }}%
    \subfloat[Voltage inputs over time]{{\includegraphics[width=0.49\textwidth]{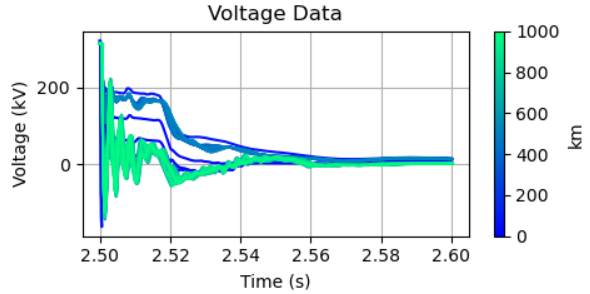} }}%
    \caption{Single ended measurements}%
    \label{fig:data}%
\end{figure*}

Fig. \ref{fig:data} shows the raw current and voltage data generated for this study. Dark color in the color bar indicates fault simulations that were closer to the beginning of the network and lighter colored lines represent simulations that were closer to the end of the network.



\section{Methodology} \label{sec:Method}
This section presents the proposed Bayesian Ridge Regression based model to predict fault locations with minimum error threshold within MTdc networks. 

Five simple pre-processing techniques are considered to process the MTdc data to predict location of faults based on Bayesian ridge regression. The data is split into k-folds to test the robustness of the proposed approach and an average of error was recorded at the end of the process.

Individual stages of processing are described below in the order in which they are applied:
\subsection*{Step-1: Downsampling (DS)}
Unbalanced datasets tend to have more observations for the response variable in one category than others and this often produces undesirable results in predicting or classifying unseen observations. To address this, the data is downsampled by a integer factor, where one-out-of-$n$ observations is selected.

\subsection*{Step-2: Principal Component Analysis (PCA)}
Principal component analysis reduces the dimensionality of the data by extracting prominent features~\cite{bishop2006pattern}. PCA is defined as the ``linear projection that minimizes the average projection cost, which is the mean squared distance between the data points and their projections''~\cite{wold1987principal}. Mathematically, minimize
\begin{equation}\label{eq:PCADef2}
J = \frac{1}{N} \sum_{i=1}^N ||x_n-\tilde{x}_n||^2
\end{equation}
\noindent where $\tilde{x}_n$ is the average of the data set. The main drawback of PCA is that the produced features no longer hold real world meaning because they are now linear combinations of other features.

\subsection*{Step-3: Low Pass Filter (LPF)}
The first two pre-processing steps allow only high frequency signal components, which could lead to data misinterpretation and is commonly known as aliasing. To address this, a finite impulse response (FIR) filter is applied as an anti-aliasing filter. It is a low pass filter, which computes every $n^{th}$ output through a decimating FIR filter computed for the $m^{th}$ sample as
\begin{equation}\label{eq:LPF}
y[m] =  \sum_{i = 0}^{k-1} x[m N-i]\bullet h[i]
\end{equation}
\noindent where $h[i]$ is the $i^{th}$ response of length $k$. 

\subsection*{Step-4: Fast Fourier Transform (FFT)}
The frequency domain components of a time-series can be extracted using Discrete Fourier Transform (DFT) \cite{He2016}. FFT is useful to perform the DFT of a sequence. FFT performs the computation of the DFT matrix as a product of sparse factors. The DFT for such a sequence can be given as (\ref{Eqn:DFT}),

\begin{equation}
    X[k] = \sum_{n= 0}^{N-1}x[n]e^{-j2\pi kn /N}
    \label{Eqn:DFT}
\end{equation}
\noindent where \emph{N} is the length of the signal. The sampling frequency of the signal is $10 kHz$, the maximum represented frequencies are half of the sampling frequency. We try to capture all the representative frequencies in that range that direct the accurate detection of a fault location. 

\subsection*{Step-5: Data Normalization}
As a last stage in pre-processing, the data is normalized to standardize it throughout the data. Normalization process is intended to eliminate any unstructured and redundant data to ensure logical data storage, thus improving the quality of the data.

Given data $X$, normalization utilizes $l^2$-norm approach as
\begin{equation}\label{eq:Norm}
\mathbf{X} = \frac{x-\mu}{|\mathbf{X}|}
\end{equation}
\noindent where $\mu$ is the mean, $\mathbf{X}$ is the $l^2$-norm, and $|\mathbf{X}| = \sqrt{\left( \sum_{i=1}^n |x_k|^2 \right)}$.

\subsection{Bayesian Ridge Regression (BRR)}\label{subsec:BayRidge}
Bayesian ridge estimates a probabilistic model of the regression problem that includes regularization parameters.

Given observations $(x_i,y_i)$, $i=1,\dots,m$ to be considered as training sequence and $x_j$ of length $k$ as a test fold split during the k-fold splitting at the beginning, the objective is to find $y_k$. In practical applications, given $(x_i,y_i)$, $i=1,\dots,n-1$ as training sequence, the objective is to predict $y_n$ for $x_n$ with smallest error.

The model assumes that the distribution of objects and labels, $(x_i,y_i)$ respectively are generated by the rule~\cite{burnaev2014efficiency}
\begin{equation}\label{eq:linear}
y_i = w \times x_i + \beta
\end{equation}
\noindent where $w$ is a random vector distribution as the Gaussian distribution parameterized by mean and covariance matrix as $N(0,(\sigma^2/a)I)$ and $\beta$ is distributed with $N(0,\sigma^2)$.

The conditional distribution for the prediction label $y_n$ for the test object $x_n$ is
$$ N(\widehat{y_n},(1+g_n)\sigma^2) $$.

 Here $g_n := x_n^'(X'X+aI)^{-1} x_n$, where $X$ is the design matrix for the training sequence. 

Now, the Bayesian prediction interval is

\begin{equation}
    \label{eq:BRRprediction}
    (B_L, B^U) := \left(\widehat{y}_n \sqrt(1+g_n)\sigma_{\epsilon/2}\widehat{y}_n + \sqrt(1+g_n)\sigma_{\epsilon/2}\right) 
\end{equation}

\section{Goodness-Of-Fit Measures}\label{subsec:GOF}
This section lists measures used to assess model performance.


\subsection{Mean Absolute Percentage Error (MAPE)}\label{sec:GOF:MAPE}
Mean absolute percentage error calculates the error as a percentage of how far off the predicted value is from the actual value
\begin{equation}
MAPE =  \frac{1}{n} \times \sum_{i=1}^{n} \bigg \lvert \frac{\widehat{x}_i - x_i}{x_i} \bigg \rvert \times 100
\end{equation}
\noindent where $\widehat{x}_i$ is the predicted value and $x_i$ is the actual value. 

\subsection{Mean Absolute Error (MAE)}\label{sec:GOF:MAE}
The performance is measured as the mean absolute (MAE) value of the prediction errors in kilometers from the target location, defined as follows,
\begin{equation}
    MAE = \frac{\sum_{i=1}^{n}\left| \widehat{x}_i-x_i\right|}{n}
\end{equation}
where $\widehat{x_i}$ is the predicted fault location, $x_i$ is the actual location and \textit{n} is the number of data points.


\subsection{Predictive Ratio Risk (PRR)}\label{sec:GOF:PRR}
The predictive ratio risk of a model is
\begin{equation}\label{eq:PRR}
    PRR = \sum_{i=k+1}^n\left(\frac{\widehat{x}_i-x_i}{\widehat{x}_i}\right)^2
\end{equation}
\noindent where the term in the denominator penalizes underestimation of the number of defects more heavily than overestimation.

\subsection{Predictive Power (PP)}\label{sec:GOF:PP}
The predictive power of a model is
\begin{equation}\label{eq:PP}
PP = \sum_{i=k+1}^n\left(\frac{\widehat{x}_i-x_i}{x_i}\right)^2
\end{equation}
\noindent where the term in the denominator penalizes overestimation of the number of defects.

\section{Results and Discussion}\label{sec:Experiment}
This section demonstrates the proposed model through examples when applied to data presented in Section~\ref{sec:Ex:ExperimentalSet}. The first example compares exhaustive combination of data pre-processing and machine learning models with the proposed approach and the second example summarizes the performance of the Bayesian ridge regression based model through comparison of goodness-of-fit measures including MAPE, MAE, PRR, and PP.

\subsection{Experimental Setup and Model Selection Validation}
To validate the selection of methods used in the proposed model, an exhaustive combination of various pre-processing methods and regression methods was applied to the data considered in Section~\ref{sec:Ex:ExperimentalSet}. A combination of five feature extraction methods, two dimensionality reduction techniques, and seven machine learning models were applied in total.

For feature extraction, low pass filter, fast Fourier transform, sklearn's Standard Scalar with mean and unit variance set to 0 and 1 respectively, $l^2$-norm based normalization method as described in Section~\ref{sec:Method}, and a square root transformation are considered. For dimensionality reduction, down sampling and principal component analysis are considered. To validate the proposed pre-processing model's efficiency in Section~\ref{sec:Method}, combinations of the seven pre-processing methods with various configurable parameter values were applied. For the techniques that do not have parameters, the list only contained two elements, True and False, corresponding to whether or not to apply the given technique. In order to adhere to laws about loss of information such as the Nyquist-Shannon sampling theorem~\cite{jerri1977shannon}, preprocessing techniques were applied in the same order every time. The parameters tested are listed below in the order that they were applied in each run of the algorithm. 
\begin{itemize}
    \item LPF: \verb|[500, 300, 250, 200, 150, 100, 50]|
    \item Down sampling Factor: \verb|[1,3,5,10,100]|
    \item Fourier Transform: \verb|True, False|
    \item Data Normalization: \verb|True, False|
    \item Principal Components: \verb|[no PCA,4,8,12,16,28]|
    \item Square Root: \verb|True, False|
    \item Standard Scalar: \verb|True, False|
\end{itemize}
Applying both standard scalar and data normalization was avoided since this would result in two levels of normalization. In addition, if the specified low-pass filter is greater than the Nyquist frequency after being downsampled, the filter would be useless because the downsampled data would not be sampled at a high enough frequency to record signals up to the filter frequency.

For each combination of preprocessing strategies, the transformed input data was passed to each of the seven machine learning regressors~\cite{bishop2006pattern, goodfellow2016deep} considered including Bayesian Ridge (BayR), Support Vector (SVR), K-Nearest Neighbors (KNNR), Decision Tree (DTree), Extreme Gradient Boosting (XGB), Gradient Boosting (GB), and Multi-Layer Perceptron (MLP). This process was done for both the voltage and current data sets. 

Models could be sensitive to traditional train-test split due to smaller sample size. To account for this this sensitivity, models were validated using k-fold cross validation with four folds. The error of each model was stored as the average of the MAE on the test sets of each of the four folds. 

\begin{figure}[]
    \centering
    \includegraphics[width=0.5\textwidth]{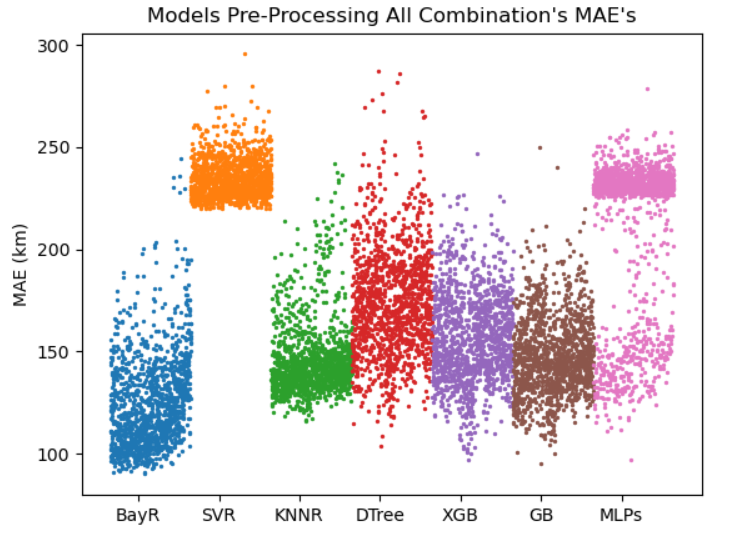}
    \caption{Pre-processing pipeline MAEs grouped by model}
    \label{fig:modelscores}
\end{figure}

Fig.~\ref{fig:modelscores} shows mean average error after testing every valid combination of preprocessing strategies and machine learning models on the input data. The results were sorted to find the models and pre-processing combinations which produced the lowest errors. Bayesian ridge regression produces the lowest errors consistently.

\subsection{Goodness-of-fit measure comparison}
Table~\ref{tbl:GOFPreProcessing} indicates decision tree as a better model based on information theoretic measures such as MAPE and MAE, while predictive measures recommend Bayesian Ridge regression as a better predictive model. BRR presents a comparatively smaller information theoretic measures with PRR and PP with at least 50\% and 70\% better predictive performance compared to all other models. 

\begin{table}[!ht]
\vspace{-3mm} 
\caption{Comparison of goodness-of-fit measures of various models}\label{tbl:GOFPreProcessing}
\centering
\begin{tabular}{|c|r|r|r|r|r|r|} \hline 
Model    & MAPE  & MAE      & PRR    & PP     \\ \hline
BayR   & 21.77 & 80.68   & \textbf{0.4748} & \textbf{0.8405} \\ \hline
SVR    & 106.1 & 219.9   & 2.607  & 63.98  \\ \hline
KNNR   & 44.80 & 111.37  & 1.135  & 1.209  \\ \hline
DTree  & \textbf{11.62} & \textbf{34.03}  & 1.361  & 4.981  \\ \hline
XGB    & 15.84 & 38.11  & 1.15   & 10.08  \\ \hline
GB      & 34.16 & 76.62  & 0.946  & 5.335  \\ \hline
MLP    & 55.20 & 117.5  & 1.502  & 14.93\\\hline
\end{tabular}
\vspace{-4mm} 
\end{table}

\begin{figure}[H]
    \centering
    \includegraphics[width=0.5\textwidth]{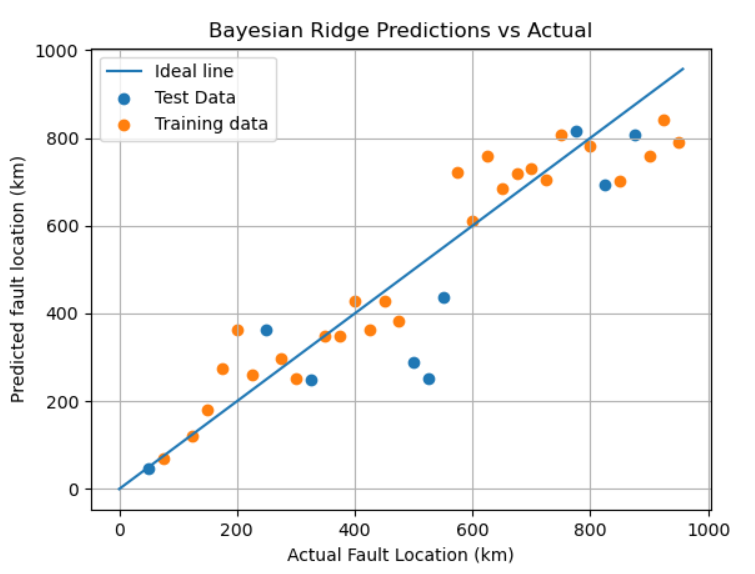}
    \caption{Bayesian Ridge regression: Predictions vs Actual}
    
    \label{fig:predictionval}
\end{figure}

Fig.~\ref{fig:predictionval} shows predictions of BRR compared to the actual fault locations for the current dataset with DS = 3, PCA = 12, LPF = 150, followed by FT and data normalization. Blue line indicates the ideal relationship between actual and estimated values.

Some other models, such as the Decision tree, scored nearly perfectly on the training data while not generalizing at all on the test data supporting results presented in Table~\ref{tbl:GOFPreProcessing}. This failure to generalize may also be seen in some of the more competitive models, such as the MLP. For these reasons, BRR is the suggested as the most desirable of the models explored here. 

\section{Conclusion} \label{sec:Conclusion}

In this paper, we proposed a novel single ended fault location technique using Bayes Regression. Post fault current and voltage measurements are recorded that are utilized to develop a probabilistic model to accurately predict fault locations in HVdc networks. We pre-process the recorded measurements using (1) \textit{Downsampling}, this helps to reduce the unbalance in the raw data that could produce biased results; (2) \textit{PCA}, helps to extract the prominent fault features from the data; (3) \textit{LPF}, this helps to reduce errors caused by aliasing; (4) \textit{FFT}, helps to extract the prominent fault frequencies essential for estimating fault location and (5) \textit{Normalization}, eliminates any unstructured or redundant data. The pre-processed data is utilized to develop a probabilistic regression model using Bayesian ridge. The performance of the prediction model is evaluated using several statistical measures discussed in Section \ref{subsec:GOF}. The obtained results demonstrates that BRR develops an effective probabilistic model of the data that provides a better predictive performance compared to other methods for accurately locating faults in multi-terminal networks.


\bibliographystyle{IEEEtran}
\bibliography{PES2022.bib}

\begin{thebibliography}{10}
\providecommand{\url}[1]{#1}
\csname url@samestyle\endcsname
\providecommand{\newblock}{\relax}
\providecommand{\bibinfo}[2]{#2}
\providecommand{\BIBentrySTDinterwordspacing}{\spaceskip=0pt\relax}
\providecommand{\BIBentryALTinterwordstretchfactor}{4}
\providecommand{\BIBentryALTinterwordspacing}{\spaceskip=\fontdimen2\font plus
\BIBentryALTinterwordstretchfactor\fontdimen3\font minus
  \fontdimen4\font\relax}
\providecommand{\BIBforeignlanguage}[2]{{%
\expandafter\ifx\csname l@#1\endcsname\relax
\typeout{** WARNING: IEEEtran.bst: No hyphenation pattern has been}%
\typeout{** loaded for the language `#1'. Using the pattern for}%
\typeout{** the default language instead.}%
\else
\language=\csname l@#1\endcsname
\fi
#2}}
\providecommand{\BIBdecl}{\relax}
\BIBdecl

\bibitem{mitra2018hvdc}
B.~Mitra, B.~Chowdhury, and M.~Manjrekar, ``Hvdc transmission for access to
  off-shore renewable energy: a review of technology and fault detection
  techniques,'' \emph{IET Renewable Power Generation}, vol.~12, no.~13, pp.
  1563--1571, 2018.

\bibitem{Kang}
S.~Kang, Y.~Wang, G.~Yang, L.~Song, and V.~Mikulovich, ``Rolling bearing fault
  diagnosis method using empirical mode decomposition and hypersphere
  multiclass support vector machine,'' \emph{Zhongguo Dianji Gongcheng
  Xuebao/Proceedings of the Chinese Society of Electrical Engineering},
  vol.~31, pp. 96--102, 05 2011.

\bibitem{Ando1985}
M.~Ando, E.~O. Schweitzer, and R.~A. Baker, ``Development and field-data
  evaluation of single-end fault locator for two-terminal hvdv transmission
  lines-part 2 : Algorithm and evaluation,'' \emph{IEEE Transactions on Power
  Apparatus and Systems}, vol. PAS-104, no.~12, pp. 3531--3537, 1985.

\bibitem{Ying-HongLin2004}
{Y. H. Lin}, {C. W. Liu}, and {C. S. Chen}, ``A new pmu-based fault
  detection/location technique for transmission lines with consideration of
  arcing fault discrimination-part ii: performance evaluation,'' \emph{IEEE
  Trans. on Power Del.}, vol.~19, no.~4, pp. 1594--1601, Oct 2004.

\bibitem{Xu2011}
Z.~{Xu} and T.~S. {Sidhu}, ``Fault location method based on single-end
  measurements for underground cables,'' \emph{IEEE Trans. on Power Del.},
  vol.~26, no.~4, pp. 2845--2854, Oct 2011.

\bibitem{Magnago1998}
F.~Magnago and A.~Abur, ``Fault location using wavelets,'' \emph{IEEE
  Transactions on Power Delivery}, vol.~13, no.~4, pp. 1475--1480, 1998.

\bibitem{Mohanty2016}
R.~{Mohanty}, U.~S.~M. {Balaji}, and A.~K. {Pradhan}, ``An accurate
  noniterative fault-location technique for low-voltage dc microgrid,''
  \emph{IEEE Trans. on Power Del.}, vol.~31, no.~2, pp. 475--481, April 2016.

\bibitem{Mitra2021}
B.~Mitra, S.~Debnath, and B.~Chowdhury, ``Fault location using the natural
  frequency of oscillation of current discharge in mtdc networks,'' \emph{IEEE
  Access}, vol.~9, pp. 49\,415--49\,423, 2021.

\bibitem{Debnath2018}
S.~{Debnath} and M.~{Chinthavali}, ``Numerical-stiffness-based simulation of
  mixed transmission systems,'' \emph{IEEE Trans. on Indus. Elect.}, vol.~65,
  no.~12, pp. 9215--9224, Dec 2018.

\bibitem{bishop2006pattern}
C.~M. Bishop, ``Pattern recognition,'' \emph{Machine learning}, vol. 128,
  no.~9, 2006.

\bibitem{wold1987principal}
S.~Wold, K.~Esbensen, and P.~Geladi, ``Principal component analysis,''
  \emph{Chemometrics and intelligent laboratory systems}, vol.~2, no. 1-3, pp.
  37--52, 1987.

\bibitem{He2016}
Z.~He, \emph{Wavelet analysis and transient signal processing applications for
  power systems}, 03 2016.

\bibitem{burnaev2014efficiency}
E.~Burnaev and V.~Vovk, ``Efficiency of conformalized ridge regression,'' in
  \emph{Conference on Learning Theory}.\hskip 1em plus 0.5em minus 0.4em\relax
  PMLR, 2014, pp. 605--622.

\bibitem{jerri1977shannon}
A.~J. Jerri, ``The shannon sampling theorem—its various extensions and
  applications: A tutorial review,'' \emph{Proceedings of the IEEE}, vol.~65,
  no.~11, pp. 1565--1596, 1977.

\bibitem{goodfellow2016deep}
I.~Goodfellow, Y.~Bengio, and A.~Courville, \emph{Deep learning}.\hskip 1em
  plus 0.5em minus 0.4em\relax MIT press, 2016.

\end{thebibliography}

\end{document}